\documentstyle[aps,prl]{revtex} 
\begin{document}

\twocolumn[\hsize\textwidth\columnwidth\hsize\csname
@twocolumnfalse\endcsname

\title{Inelastic lifetimes of hot electrons in real metals}
\author{I. Campillo$^{1}$, J. M. Pitarke$^{1,4}$, A. Rubio$^{2}$,
E. Zarate$^{3}$, and P. M. Echenique$^{3,4}$}
\address{$^1$ Materia Kondentsatuaren Fisika Saila, Zientzi Fakultatea, 
Euskal Herriko Unibertsitatea,\\ 644 Posta kutxatila, 48080 Bilbo, Basque 
Country,
Spain\\
$^2$Departamento de F{\'{\i}}sica Te\'orica, Universidad de Valladolid,
47011 Valladolid, Spain\\
$^3$ Materialen Fisika Saila, Kimika Fakultatea, Euskal Herriko 
Unibertsitatea,\\
1072 Posta kutxatila, 20080 Donostia, Basque Country, Spain\\
$^4$Donostia International Physics Center (DIPC) and Centro Mixto CSIC-UPV/EHU, Spain}

\date\today

\maketitle

\begin{abstract}
We report a first-principles description of inelastic lifetimes of excited 
electrons in real Cu and Al, which we compute, within the GW approximation of
many-body theory, from the knowledge of the self-energy of the excited
quasiparticle. Our full band-structure calculations indicate that actual lifetimes
are the result of a delicate balance between localization, density of states,
screening, and Fermi-surface topology. A major contribution from $d$-electrons
participating in the screening of electron-electron interactions yields lifetimes of
excited electrons in copper that are larger than those of electrons in a
free-electron gas with the electron density equal to that of valence ($4s^1$)
electrons. In aluminum, a simple metal with no $d$-bands, splitting of the band
structure over the Fermi level results in electron lifetimes that are smaller than
those of electrons in a free-electron gas.
\end{abstract}
]

\narrowtext

Electron dynamics in metals are well-known to play an important role in a
variety of physical and chemical phenomena, and low-energy excited electrons have been
used as new probes of many-body decay mechanisms and chemical
reactivity\cite{Petek1}. Recently, the advent of time-resolved two-photon
photoemission (TR-2PPE)\cite{Fauster} has made possible to provide direct
measurements of the lifetime of these so-called hot electrons in
copper\cite{exp1,exp2,exp3,exp4,exp6,Goldmann}, other noble and
transition metals\cite{Knoesel}, ferromagnetic solids\cite{exp5}, and high
$T_c$ superconductors\cite{Petek2}. Also, ballistic electron emission spectroscopy
(BEES) has shown to be capable of determining hot-electron relaxation times
in solid materials\cite{fj}.

An evaluation of the inelastic lifetime of excited electrons in the vicinity of
the Fermi surface was first reported by Quinn and Ferrell\cite{QF}, within a
many-body free-electron description of the solid, showing that it is inversely
proportional to the square of the energy of the quasiparticle measured with
respect to the Fermi level. Since then, several free-electron calculations of
electron-electron scattering rates have been performed, within the random-phase
approximation (RPA)\cite{Ritchie,Shelton} and with inclusion of exchange
and correlation effects\cite{Ashley,Penn}. Band structure effects were
discussed by Quinn\cite{Quinn2} and Adler\cite{Adler}, and statistical
approximations were applied by Tung {\it et al}\,\cite{Tung} and by
Penn\cite{Penn2}. Nevertheless, there was no first-principles calculation of
hot-electron lifetimes in real solids, and further theoretical work was needed for
the interpretation of existing lifetime measurements and, in particular, to
quantitatively account for the interplay between band-structure and many-body effects
on electron relaxation processes.

In this Letter we report results of a full {\it ab initio} evaluation of relaxation
lifetimes of excited electrons in real solids.
First, we expand the one-electron Bloch states in
a plane-wave basis\cite{cut}, and solve the Kohn-Sham equation of density-functional
theory (DFT)\cite{Kohn} by invoking
the local-density approximation (LDA) for exchange and correlation\cite{Ceperley}. The
electron-ion interaction is described by means of a non-local, norm-conserving
ionic pseudopotential\cite{Troullier}, and we use the one-electron Bloch states 
to
evaluate the screened Coulomb interaction within a well-defined many-body 
framework,
the RPA\cite{Fetter}. We finally evaluate the lifetime
from the knowledge of the imaginary part of the electron self-energy of the 
excited
quasiparticle, which we compute within the so-called GW approximation of 
many-body
theory\cite{Hedin}.  

Let us consider an inhomogeneous electron system.
The damping rate of an excited electron in
the state $\phi_0({\bf r})$ with energy $E$ is obtained as (we use atomic
units throughout, i. e., $e^2=\hbar=m_e=1$)
\begin{equation}\label{eq1}
\tau^{-1}=-{2}\int{\rm d}{{\bf r}}\int{\rm d}{{\bf
r}'}\phi_{0}^*({\bf r}){\rm Im}\,\Sigma({\bf r},{\bf r}';E)
\phi_{0}({\bf r}'),
\end{equation}
where $\Sigma({\bf r},{\bf r}';E)$ represents the electron self-energy.
In the so-called GW approximation, only the first term of the expansion of the
self-energy in the screened interaction is considered, and after replacing the
Green function ($G$) by the zero order approximation ($G^0$), one finds
\begin{equation}
{\rm Im}\,\Sigma({\bf r},{\bf r}';E)=\sum_f\phi_f^*({\bf r}')
{\rm Im}\,W({\bf r},{\bf r}';\omega)\phi_f({\bf r}),
\end{equation}
where $\omega=E-E_f$ represents the energy transfer, the sum is extended over a 
complete
set of final states $\phi_f({\bf r})$ with energy $E_f$ ($E_F\le E_f\le
E$), $E_F$ is the Fermi 
energy, and $W({\bf r},{\bf r}';\omega)$ is the screened Coulomb interaction:
\begin{equation}\label{eqint}
W({\bf r},{\bf r}';\omega)=\int{\rm d}{\bf r}''\epsilon^{-1}({\bf r},{\bf
r}'',\omega)v({\bf r}''-{\bf r}').
\end{equation}
Here, $v({\bf r}-{\bf r}')$ represents the bare Coulomb interaction, and
$\epsilon^{-1}({\bf r},{\bf r}',\omega)$ is the inverse dielectric function of the
solid, which we evaluate within RPA\cite{Quong}.

We introduce Fourier expansions appropriate for periodic crystals, and find
\begin{eqnarray}\label{eq7}
\tau^{-1}={1\over\pi^2}\sum_f
\int_{\rm BZ}{{\rm d}{\bf q}}\sum_{\bf G}\sum_{{\bf G}'}&&
{B_{0f}^*({\bf q}+{\bf G})B_{0f}({\bf q}+{\bf G}')\over\left|{\bf q}+{\bf 
G}\right|^2}\cr
&&\times{\rm Im}\left[-\epsilon_{{\bf G},{\bf G}'}^{-1}({\bf
q},\omega)\right],
\end{eqnarray}
where
\begin{equation}\label{eq8}
B_{0f}({\bf q})=\int{\rm d}^3{\bf r}\phi_0^{\ast}({\bf r}){\rm e}^{{\rm
i}{\bf q}\cdot{\bf r}}\phi_f({\bf r}),
\end{equation}
and where $\epsilon^{-1}_{{\bf G},{\bf G}'}({\bf q},\omega)$ represent Fourier
coefficients of the inverse dielectric function of the crystal. ${\bf G}$ and
${\bf G}'$ are reciprocal lattice vectors, and the integration over
$\bf q$ is extended over the first Brillouin zone (BZ).

In particular, if couplings of the wave vector ${\bf q}+{\bf G}$ to wave vectors 
${\bf
q}+{\bf G}'$ with ${\bf G}\neq{\bf G}'$, i. e., the so-called crystalline 
local-field
effects are neglected, one can write:
\begin{equation}\label{eq7p}
\tau^{-1}={1\over\pi^2}\sum_f
\int_{\rm BZ}{{\rm d}{\bf q}}\sum_{\bf G}
{\left|B_{0f}({\bf q}+{\bf G})\right|^2\over\left|{\bf q}+{\bf G}\right|^2}
{{\rm Im}\left[\epsilon_{{\bf G},{\bf G}}({\bf q},\omega)\right]\over 
|\epsilon_{{\bf
G},{\bf G}}({\bf q},\omega)|^2}.
\end{equation}
If all one-electron Bloch states entering both the matrix elements $B_{0f}({\bf 
q}+{\bf
G})$ and the dielectric function $\epsilon_{{\bf G},{\bf G}}({\bf q},\omega)$ 
are
represented by plane waves, then Eq. (\ref{eq7p}) exactly
coincides with the GW formula for the scattering rate of
excited electrons in a free-electron gas (FEG), as obtained by Quinn and
Ferrell\cite{QF} and by Ritchie\cite{Ritchie}. In the case of electrons with 
energy
very near the Fermi energy ($E\approx E_F$) the phase space available for real
transitions is simply $E-E_F$, which yields the well-known $(E-E_F)^2$ scaling of
the scattering rate. In the high-density limit ($r_s<<1$)\cite{rs}, one
finds\cite{QF}
\begin{equation}\label{eqqf}
\tau_{QF}=263\,r_s^{-5/2}(E-E_F)^{-2}\, {\rm eV}^2\,{\rm fs}.
\end{equation}

The hot-electron decay in real solids depends on both the wave vector
${\bf k}$
and the band index $n$ of the initial Bloch state $\phi_0({\bf r})={\rm
e}^{{\rm
i}{\bf k}\cdot{\bf r}}u_{{\bf k},n}({\bf r})$. We have evaluated hot-electron
lifetimes along various directions of the wave-vector\cite{Igor2}, and have found that
scattering rates of low-energy electrons are strongly directional
dependent. Since measurements of hot-electron lifetimes have been
reported as a function of energy, we here focus on the evaluation of $\tau^{-1}(E)$,
which we obtain by averaging $\tau^{-1}({\bf k},n)$ over all wave vectors and bands
lying in the irreducible element of the Brillouin zone (IBZ) with the same energy. The
results presented below have been found to be well converged for all hot-electron
energies under study ($E-E_F=1.0-3.5\,{\rm eV}$), and they all have been performed
with inclusion of conduction bands up to a maximum energy of
$\sim 25\,{\rm eV}$ above the Fermi level. The sampling of the BZ required for
the evaluation of both the dielectric matrix and the hot-electron
decay rate
of Eqs. (\ref{eq7}) and (\ref{eq7p}) has been performed on
$16\times 16\times 16$ Monkhorst-Pack meshes\cite{Monk}. The sums in Eqs.
(\ref{eq7}) and (\ref{eq7p}) have been extended over
$15\,{\bf G}$ vectors of the reciprocal lattice, the magnitude of the maximum
momentum transfer ${\bf q}+{\bf G}$ being well over the upper limit of $\sim
2\,q_F$ ($q_F$ is the Fermi momentum).

Our {\it ab initio} calculation of the average lifetime $\tau(E)$ of hot
electrons in Cu, as obtained from Eq. (\ref{eq7}) with full inclusion of crystalline
local-field effects, is presented in Fig. 1 by solid
circles. The lifetime of  hot
electrons in a FEG with the electron density equal to that of 
valence
($4s^1$) electrons in copper ($r_s=2.67$) is exhibited in the same figure, by a 
solid
line. Both calculations have been carried out within one and the same many-body
framework, the RPA; thus, the ratio between our calculated {\it ab initio} 
lifetimes and
the corresponding FEG calculations unambiguously establishes the impact of the band
structure of  the
crystal on the hot-electron decay. Our {\it ab initio} calculations indicate 
that the
lifetime of hot electrons in copper is, within RPA, larger than that of
electrons in a FEG with $r_s=2.67$, this enhancement varying from a factor of
$\sim 2.5$ near the Fermi level ($E-E_F=1.0\,{\rm eV}$) to a factor of
$\sim 1.5$ for $E-E_F=3.5\,{\rm eV}$. We have also performed calculations of the 
lifetime
of hot electrons in copper by just keeping the $4s^1$ Bloch states as valence 
electrons
in the pseudopotential generation. The result of this calculation nearly 
coincides with
the FEG calculations, showing the key role that $d$-states play in the 
electron-decay mechanism.

First of all, we focus on the role that both localization and density of states
(DOS) available  for real excitations play in the
hot-electron lifetime. Hence, we neglect crystalline local-field effects and evaluate
hot-electron lifetimes from Eq. (\ref{eq7p}) by replacing the electron initial and
final states in $\left|B_{0f}({\bf
q}+{\bf G})\right|^2$ by plane waves, and the dielectric function in 
$\left|\epsilon_{{\bf G},{\bf G}}({\bf q},\omega)\right|^{-2}$ by that of a FEG with
$r_s=2.67$. The result we obtain, with full inclusion of the band structure of the
crystal in the evaluation of ${\rm Im}\left[\epsilon_{{\bf G},{\bf G}}({\bf
q},\omega)\right]$, is  represented in Fig. 1 by open circles. Since the states just
below the Fermi level, which are available for real transitions, have a
small but significant $d$-component, they are more localized than pure $sp$-states.
Hence, their overlap with states over the Fermi level is smaller than in
the case of free-electron states, and localization results in lifetimes of electrons
with $E-E_F<2\,{\rm eV}$ that are slightly larger  than predicted within the FEG model
of the metal (solid line). At larger energies  this band structure calculation
predicts a lower lifetime than within the FEG model, due to opening of the $d$-band
scattering channel dominating the DOS with energies from  $\sim 2.0\,{\rm eV}$ below
the Fermi level.
 
While the excitation of $d$-electrons diminishes the lifetime of electrons with
energies $E-E_F>2\,{\rm eV}$, $d$-electrons also give rise to additional screening,
thus increasing the lifetime of {\it all} electrons above the Fermi level. That
this is the case is obvious from our band-structure calculation exhibited by full
triangles in Fig. 1. This calculation is the result obtained from  Eq. (\ref{eq7p}) by
still replacing  hot-electron initial and final states in $\left|B_{0f}({\bf q}+{\bf 
G})\right|^2$ by
plane waves (plane-wave calculation), but including the full band structure of 
the crystal
in the evaluation of both
${\rm Im}\left[\epsilon_{{\bf G},{\bf G}}({\bf q},\omega)\right]$ and
$\left|\epsilon_{{\bf G},{\bf G}}({\bf q},\omega)\right|^{-2}$. The effect of
virtual interband transitions giving rise to additional screening is to 
increase, for the energies under study, the lifetime by a factor of $\approx 3$, in 
qualitative
agreement with the approximate prediction of Quinn\cite{Quinn2}.

Finally, we investigate band structure effects on hot-electron energies
and wave functions. We have performed band-structure calculations of Eq.
(\ref{eq7})
with and without (see also Eq. (\ref{eq7p})) the inclusion of crystalline local field
corrections, and we have found that these corrections are negligible for 
$E-E_F>1.5\,{\rm
eV}$, while for energies very near the Fermi level neglection of these 
corrections
results in an overestimation of the lifetime of less than $5\%$. Therefore,
differences between our full (solid circles) and plane-wave (solid
triangles) calculations come from the sensitivity of hot-electron initial and
final states on the band structure of the crystal. When the hot-electron energy 
is well
above the Fermi level, these states are very nearly plane wave states for most 
of the
orientations of the wave vector, and the lifetime is well described by 
plane-wave
calculations (solid circles and triangles nearly coincide for $E-E_F>2.5\,{\rm 
eV}$).
However, in the case of hot-electron energies near the Fermi level, initial and 
final
states strongly depend on the orientation of the wave vector and on the shape of 
the Fermi
surface. While the lifetime of hot electrons with the wave vector along the
necks of the Fermi surface, in the $\Gamma L$ direction, is found to be longer than the
averaged lifetime by up to $80\%$, flattening of the Fermi surface along the $\Gamma K$
direction is found to increase the hot-electron decay rate by up to $10\%$ (see also
Ref.\onlinecite{Adler}). Since for most orientations the Fermi surface is flattened,
Fermi surface shape effects tend to decrease the average inelastic lifetime. An
opposite behaviour occurs for hole-states with a strong $d$-character below the Fermi 
level, localization of these states strongly increasing the  hole-lifetime\cite{Igor2}.

Our band-structure calculation of hot-electron lifetimes in Al, as obtained from Eq.
(\ref{eq7}), is exhibited in the inset of Fig. 1 by solid circles, together with
lifetimes of hot electrons in a FEG with
$r_s=2.07$ (solid line). Since aluminum is a simple metal with no $d$-bands, ${\rm
Im}\left[-\epsilon_{{\bf G},{\bf G}'}^{-1}({\bf  q},\omega)\right]$ is well 
described
within a free-electron model, and band structure effects only enter through the
sensitivity of hot-electron initial and final wave functions on the band structure of
the crystal. Due to splitting of the band structure over the Fermi level new decay
channels are opened, and band structure effects now tend to
decrease the hot-electron lifetime by a factor that varies from $\sim 0.65$ near the
Fermi level ($E-E_F=1\,{\rm eV}$) to a factor of
$\sim 0.75$ for
$E-E_F=3\,{\rm eV}$. 

Scaled lifetimes of hot electrons in Cu, as determined from
most recent TR-2PPE experiments\cite{exp3,exp4,exp6}, are represented in Fig. 2,
together with our calculated lifetimes of hot electrons in the real crystal (solid
circles) and in a FEG with $r_s=2.67$ either within the full RPA (solid line) or with
use of Eq. (\ref{eqqf}) (dashed line). Though there are large discrepancies among
results obtained in different laboratories, most experiments give lifetimes that are
considerably longer than predicted within a free-electron description of the metal. At
$E-E_F<2\,{\rm eV}$, our calculations are close to lifetimes recently measured by
Knoesel {\it et al} in the very-low energy range\cite{exp6}. At larger electron
energies, good agreement between our band-structure calculations and
experiment is obtained for Cu(110), where no band gap exists in the ${\bf
k}_\parallel=0$ direction. However, one must be cautious in the interpretation of
TR-2PPE lifetime measurements, since they may be sensitive to electron
transport away from the probed surface and also to the presence of the hole left
behind in the photoexcitation process. In the case of injected electrons in BEES
experiments no hole is present. A careful analysis of electron-electron mean free
paths in these experiments\cite{fj} has shown a
$(E-E_F)^{-2}$ dependence of hot-electron lifetimes in the noble metals, with an
overall enhancement with respect to those predicted by Eq. (\ref{eqqf}) by a factor of
$\sim 2$, in agreement with our band-structure calculations. We note that our
calculated lifetimes (solid circles) approximately scale as $(E-E_F)^{-2}$, as a
result of two competing effects. As the energy increases hot-electron lifetimes in a
FEG (solid line) are known to be larger than those predicted by Eq. (\ref{eqqf})
(dashed line)\cite{review}, and this enhancement of the lifetime is nearly compensated
by the reduction, at energies well over the Fermi level, of band structure effects.

In conclusion, we have performed full RPA band-structure calculations of hot-electron
inelastic lifetimes in real solids, and have demonstrated that decay rates of excited
electrons strongly depend on details of the electronic band structure. In the case of
Cu, a subtle competition between localization, density of states, screening, and
Fermi-surface topology results in hot-electron lifetimes that are larger than those
of electrons in a FEG, and good agreement is obtained, for $E-E_F>2\,{\rm eV}$, with
observed lifetimes in Cu(110). For Al, interband transitions over the Fermi
level yield hot-electron lifetimes that are smaller than those of electrons in a FEG. 

We acknowledge partial support by the University of the Basque Country, the
Basque Unibertsitate eta Ikerketa Saila, and the Spanish Ministerio de
Educaci\'on y Cultura.

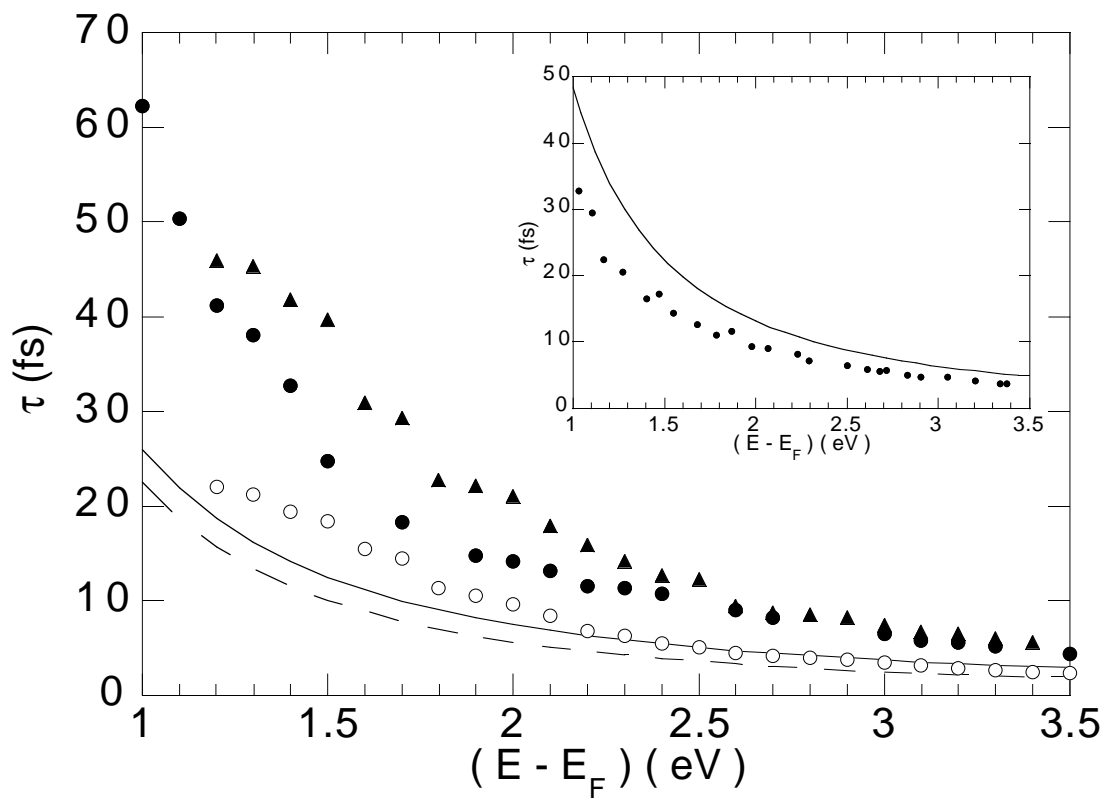
\begin{figure}
\caption{Hot-electron lifetimes in Cu. Solid circles represent our full {\it ab 
initio} calculation of $\tau(E)$, as obtained after averaging $\tau({\bf k},n)$
of Eq. (\ref{eq7}) over wave vectors and over the band structure for each ${\bf k}$. 
Solid and dashed lines represent the lifetime of hot electrons in a 
FEG with $r_s=2.67$, as obtained within the full RPA (solid line) 
and from Eq. (\ref{eqqf}) (dashed line). Open circles represent the
result obtained from Eq. (\ref{eq7p}) by replacing hot-electron initial and final
states in $\left|B_{0f}({\bf
q}+{\bf G})\right|^2$ by plane waves and the dielectric function in
$\left|\epsilon_{{\bf
G},{\bf G}}({\bf q},\omega)\right|^{-2}$ by that of a FEG with $r_s=2.67$,
but with full inclusion of the band structure in the calculation of ${\rm 
Im}\left[\epsilon_{{\bf G},{\bf G}}({\bf
q},\omega)\right]$. Full triangles represent the result obtained from Eq. (\ref{eq7p})
by  replacing
hot-electron initial and final
states in $\left|B_{0f}({\bf
q}+{\bf G})\right|^2$ by plane waves, but with full inclusion of the band 
structure in the evaluation of both ${\rm Im}\left[\epsilon_{{\bf
G},{\bf G}}({\bf q},\omega)\right]$ and $\left|\epsilon_{{\bf
G},{\bf G}}({\bf q},\omega)\right|^{-2}$. The inset exhibits hot-electron
lifetimes in Al. Solid circles represent our {\it ab initio}
calculations, and the solid line represents lifetimes of hot electrons in a FEG
with $r_s=2.07$, as obtained within the full RPA.}
\end{figure}

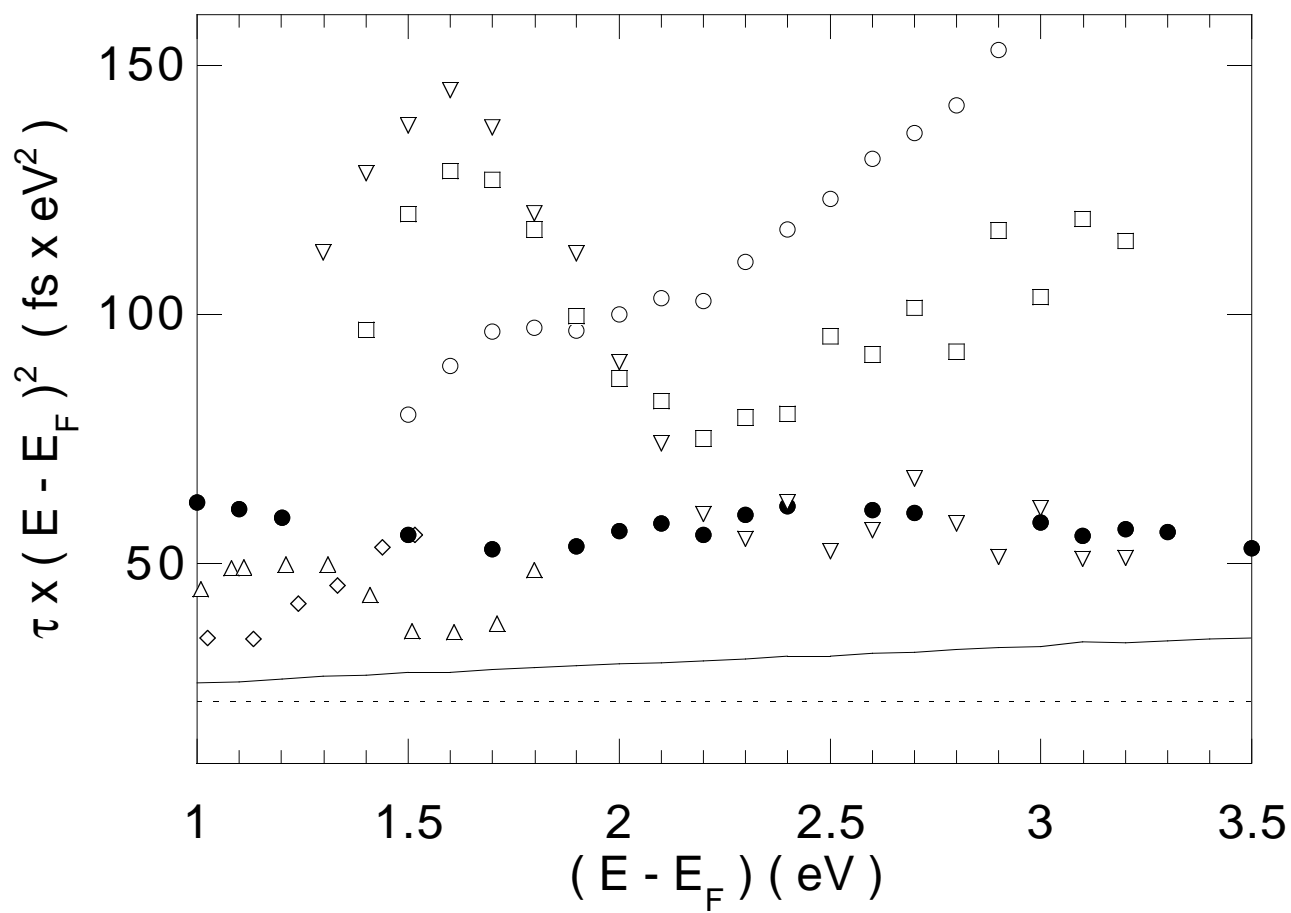
\begin{figure}
\caption[]{Scaled lifetimes, $\tau\times(E-E_F)^2$, of hot electrons in Cu, as
obtained from Eq. (\ref{eq7}) with full inclusion of crystalline local-field effects
(solid circles). Solid and dashed lines represent scaled lifetimes of hot electrons in
a FEG with $r_s=2.67$, as obtained within the full RPA (solid line)  and from Eq.
(\ref{eqqf}) (dashed line). Experimental lifetimes are taken from
Ref.{\onlinecite{exp3}} (Cu[100]: squares, Cu[110]: inverted triangles,
Cu[111]: open circles), from Ref.{\onlinecite{exp4}} with
$\omega_{fot}=1.63\,{\rm eV}$ (diamonds), and from Ref.{\cite{exp6}} (triangles).}
\end{figure}


\begin{references}
\bibitem{cut} Kinetic-energy cutoffs of $75\,{\rm Ry}$ and $12\,{\rm Ry}$ are
introduced for Cu and Al, respectively. In the case of copper, all $4s^1$ and
$3d^{10}$ Bloch states are kept as valence electrons in the pseudopotential
generation.
\bibitem{rs} The so-called electron-density parameter $r_s$ is defined by
the relation $1/n_0=(4/3)\pi(r_s\,a_0)^3$, $n_0$ being the average electron 
density, and $a_0$, the Bohr radius.
\end{references}
\end{document}